\documentclass [12pt]{article}
\usepackage {amssymb}
\usepackage {amsmath}
\usepackage{graphicx}
\usepackage {longtable}

\title{Statistical properties of telephone communication network}

\date{}

\author{ \textbf{V.M.Danilevskiy}$^{1}$,
\textbf{V.V.Yanovsky}$^{1,2}$}
\begin{document}

 \maketitle
$^{1}$\textit{V. N. Karazin Kharkiv National University, 4 Svobody Sq., Kharkiv 61022, Ukraine}

$^{2}$\textit{Institute for Single Crystals, NAS Ukraine, 60 Nauky Ave., Kharkov 61001, Ukraine}

\abstract{The directed network of telephone subscribers is considered in the article. It can be described as a dynamic network with vertices that correspond to the subscribers of the telephone network and emerging directional edges that correspond to the connections between the respective subscribers. The position of the edge and its direction is determined by the incoming and outgoing calls from the corresponding vertices. The subject of the article is the statistical properties of the connections of a certain subset of telephone network subscribers. Such connections are dynamic in nature due to their appearance and disappearance. The number of outgoing (or incoming) connections occurred during a day at a selected vertex is used as the main characteristic. The distribution density of the number of outgoing (or incoming) connections (or calls) of such a network has been analyzed using the experimental data. It has been shown that such a distribution density over the number of calls obeys the lognormal distribution density, which depends on the two parameters. The values of two parameters, namely the mean value and the variance, determining the lognormal distribution density are established. The reasons for the appearance of a lognormal distribution density over the number of incoming (or outgoing) connections have been discussed. The statistical properties of other groups of subscribers have been considered as well. In particular, the group that makes a large number of outgoing calls to various subscribers of the telephone network has been selected for a separate study. The members of this group, who create and distribute spam can be called spammers. It has been shown that these groups, spammers for example, also obeys the lognormal distribution density over the number of calls but they are characterized by the different mean value and variance.}

\section{Introduction}

The study of complex networks and their properties has been intensively developing recently. The study was started by Euler, who solved the problem of seven bridges and gave rise to graph theory. One of the well-known achievements of graph theory is the proof of the four-color theorem proved by Appel and Haken \cite{1v}. The next step to the theory of complex networks was the appearance of random graphs in the works of Erdos and Renyi \cite{2v}. The increasing number of examples of non-traditional complex networks has led to intensive research. Significant results were obtained in statistical physics, a review of which can be found in \cite{3v,4v,5v,6v}. The area of complex networks has penetrated biology \cite{7v,8v,9v}, economics and social systems \cite{10v,11v,12v,13v}, ecology \cite{14v,15v,16v} and many other areas.

In \cite{16a, 17a}, studying the properties of the Internet network, the degree distribution of the network was found to be close to the power law. Similar behavior was found in \cite{19v,20v}. Apparently, distribution close to power-law distribution is widespread in various networks. In \cite{17v,18v}, a two parametric model of random graphs was proposed, which demonstrates such a distribution of degrees of vertices. In \cite{21v}, it was shown that the set of degrees of the so-called call graph is well approximated by a power distribution. Call graphs are graphs processed by some subsets of telephony operators over a given period. In \cite{18v}, small differences from power-law behavior were also noted.

In this paper, we consider the properties of a telephone network close to call graphs. The vertices of the graph correspond to certain subscribers of the telephone network, i.e. a subset of telephone subscribers. Each vertex makes a certain number of calls and receives a different number of calls for a certain time interval. As usual, we will distinguish between incoming and outgoing calls. Then a dynamic graph appears with certain numbers of incoming and outgoing edges. Edges appear and disappear. It can be considered as a complex dynamic network of calls. The main question comes down to the distribution density of the number of outgoing (or incoming) calls for a certain time interval. In other words, what is the probability of making n calls by some subscriber for a certain time interval (for example, per day). For the selected vertex, a number of incoming calls in one day correspond to the number of edges included in it or an integral degree of a vertex for a given period (for instance, per day). This corresponds to the distribution of the number of vertices from the values of their degrees. In this work, such a distribution is constructed according to experimental data. It is shown that it is the lognormal distribution density. The properties of the density distribution between different groups of subscribers, including spammers, are discussed. Besides, a possible reason for the appearance of such distribution density is given.

\section{The distribution function of phone calls}
\begin{figure}
  \centering
  \includegraphics[width=5 cm]{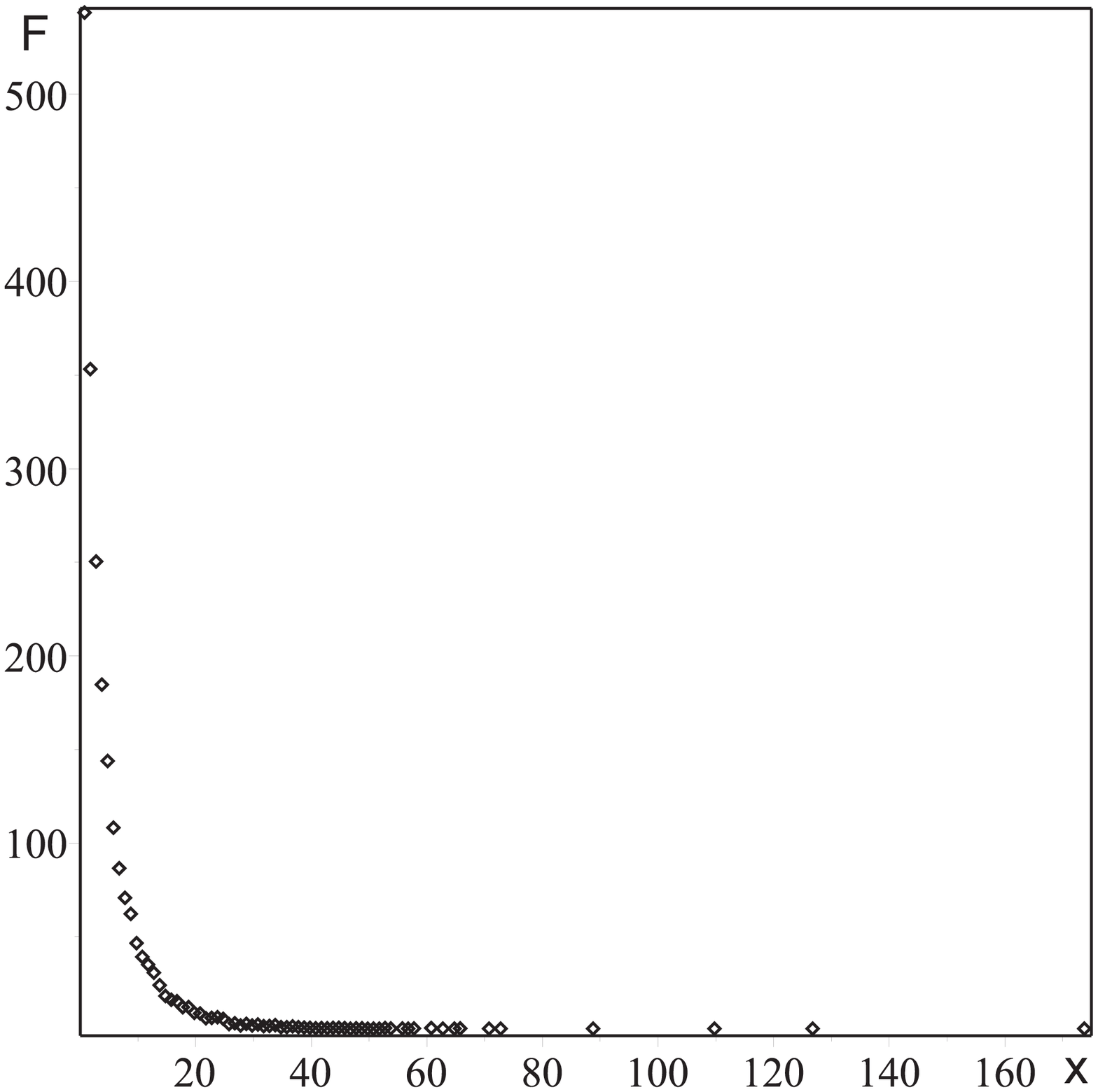}
  \includegraphics[width=5 cm]{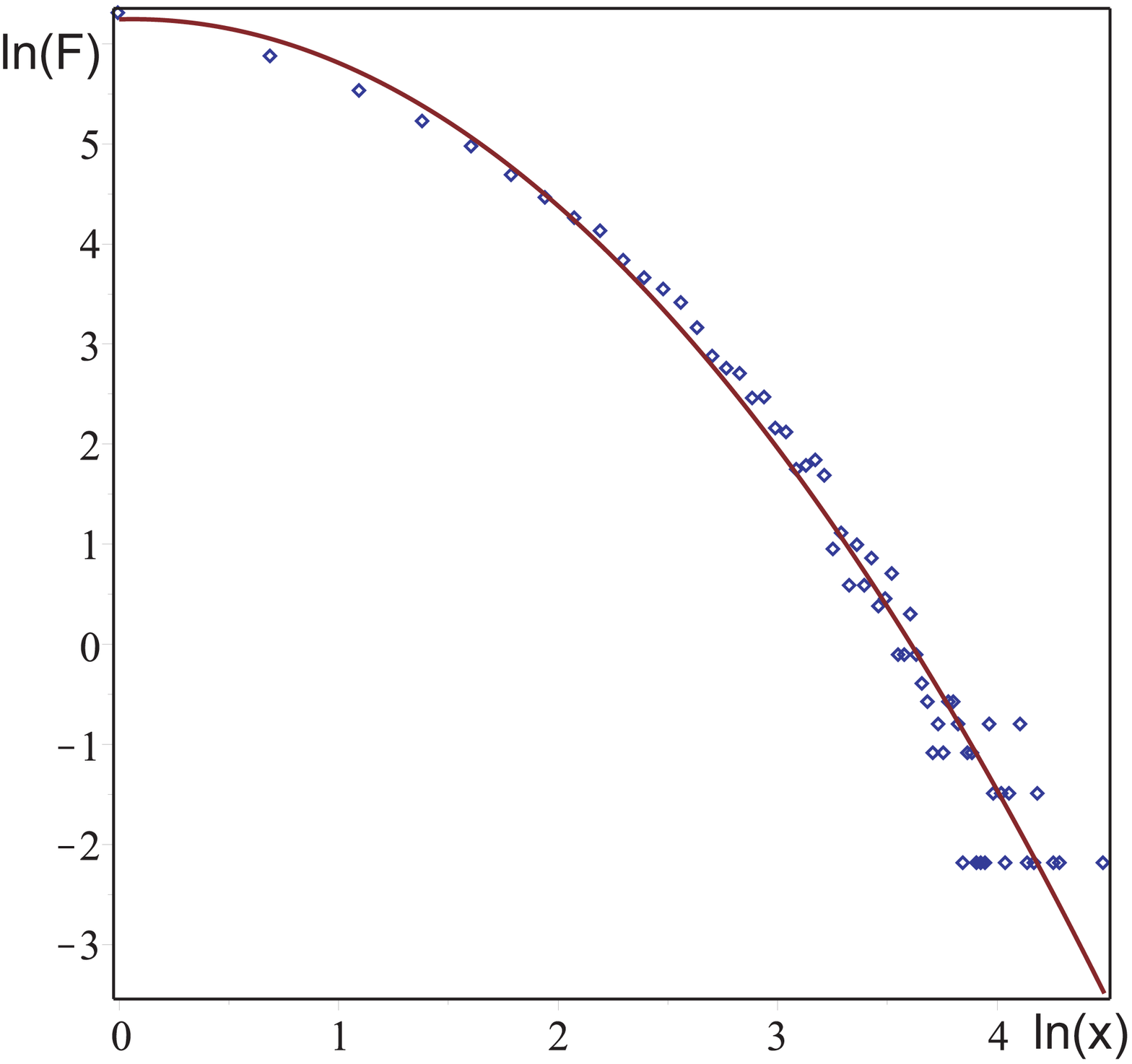}
  \includegraphics[width=5 cm]{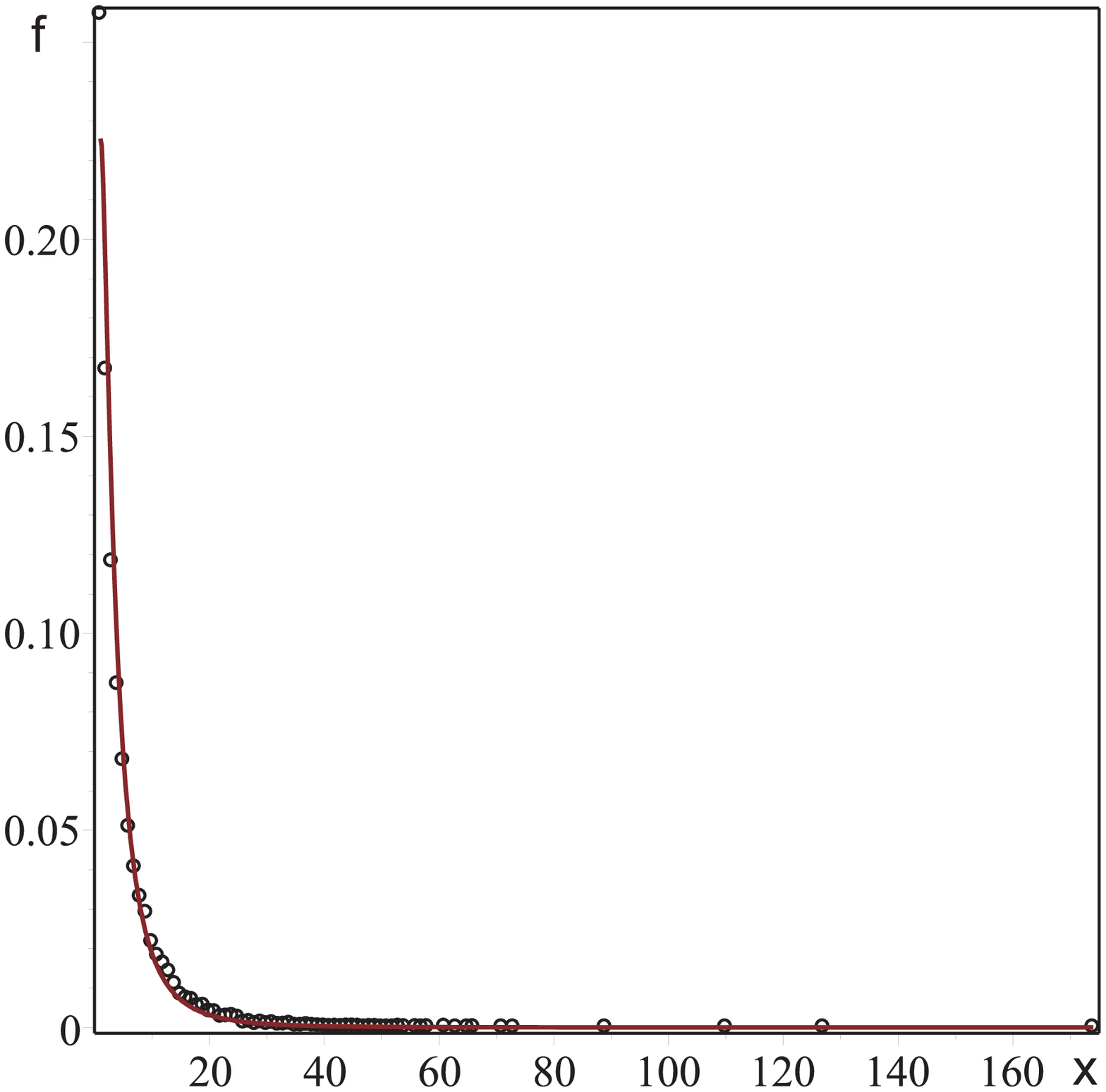}
  \caption{Distribution function $F$ of outgoing calls per day(left). $x$ -- number of outgoing calls per day, $F(x)$ -- number of subscribers, who made x calls over the day. Averaging was carried out over 9 days. Data in logarithmic coordinates (middle). The normalized distribution density of outgoing calls is shown by circles (right). The continuous curve corresponds to the lognormal distribution density with the obtained parameters $\sigma$ and $\mu$ from the approximating dependence. }\label{fg1}
\end{figure}

Let's pick a subset of US telephone network subscribers. The size of a subset is $N_0 \approx 4200$ subscribers. Consider their activity on incoming and outgoing calls for 1 day. We start with a discussion of experimental data on outgoing telephone network calls. To do this, using data on calls for the day, we plot a histogram of the distribution of outgoing calls for this day. Such a histogram can be smoothed using averaging of such histograms obtained from data for different days. The work uses data that was observed for 9 days (from 2017-11-01 to 2017-11-9 inclusive). A histogram in Fig.\ref{fg1} illustrates the experimental number of subscribers who called in a single day a certain number of times (left). The histogram was obtained by averaging over 9 days. The total number of outgoing calls of this subset, according to the experimental data of the histogram, is  $N_{out} \approx 2110$. In other words, less than half of the subscribers of the selected subset made an outgoing call during the day. To establish the dependence of the density of the distribution function, we construct this histogram on a logarithmic scale (middle graph in Fig.\ref{fg1}). It is easy to notice that the arrangement of experimental values is specific for the parabolic curve. The quadratic approximation of these values by the least squares method leads to a function  $\ln (y) = 6.24 + 0.62 \ln (x) - 0.50 \ln (x)^2$.  Such dependence is specific for a lognormal distribution.

Thus, using the experimental data, we obtain the form of the experimental distribution density of outgoing calls, normalizing the histogram values to the total number of calls. The corresponding dependence is shown in Fig.\ref{fg1}(right).

The lognormal distribution density function has the form \cite{1s},\cite{2s}
\begin{equation}\label{e1}
  f(x) = \left \{
  \begin{array}{rl}
           \frac{e^{-\frac{(\ln x -\mu)}{2 \sigma^2}}}{x \sigma \sqrt{2 \pi}}& \mbox{if} \quad x>0 \\
           0 & \mbox{if} \quad x \leq 0
         \end{array}\right.
\end{equation}
where $\sigma$ and $\mu$ are two parameters that determine the lognormal distribution density.

So the mathematical expectation $\mu^{\ast}=e^{\mu +\frac{\sigma^2}{2}}$ and variance $\sigma^{\ast 2}=e^{2\mu +\sigma^2}(e^{\sigma^2}-1)$ of the lognormal distribution are expressed through these parameters. If we designate the coefficients included in the approximating function as $\ln(y)=c+b \cdot \ln(x)- a \cdot \ln(x)^2$, then we can set the parameters of the lognormal distribution density. Really easy to derive $\mu =\frac{b+1}{2 a}$ and $\sigma^2 =\frac{1}{2 a}$. Using the experimental values $a$, $b$ we obtain $\mu \approx 1.1$ and $\sigma^{2} \approx 1.0$. It should be emphasized that the obtained parameter values are close to 1. Then the mathematical expectation of the lognormal distribution $\mu^{\ast}\approx 8$, and the variance $\sigma^{\ast 2} \approx 40$.

Therefore, the distribution density by the number of outgoing calls coincides with the lognormal distribution. According to experimental data the parameters of this distribution are $\mu \approx 1.1$ and  $\sigma^{2} \approx 1$.

\begin{figure}
  \centering
  \includegraphics[width=5 cm]{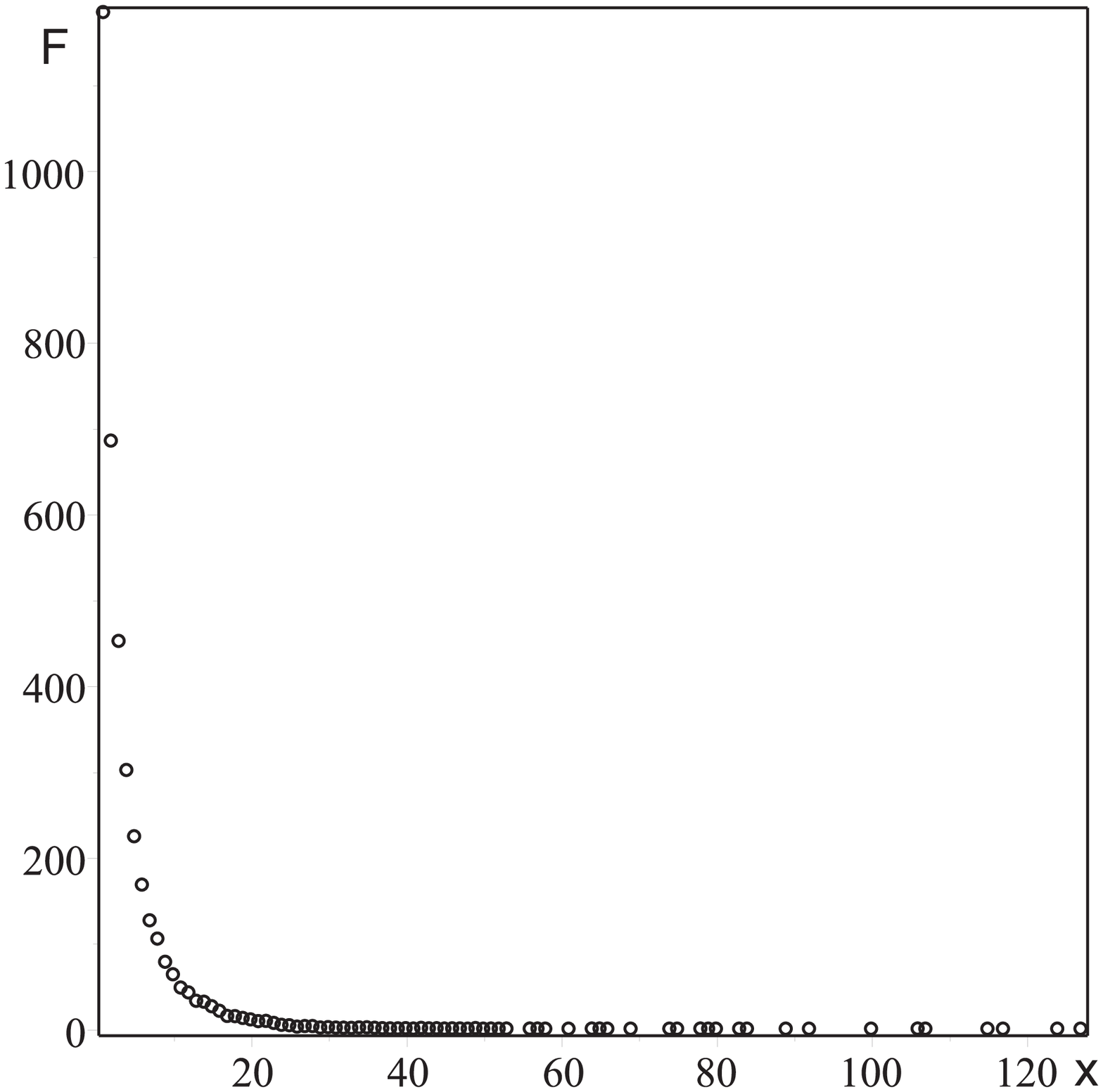}
   \includegraphics[width=5 cm]{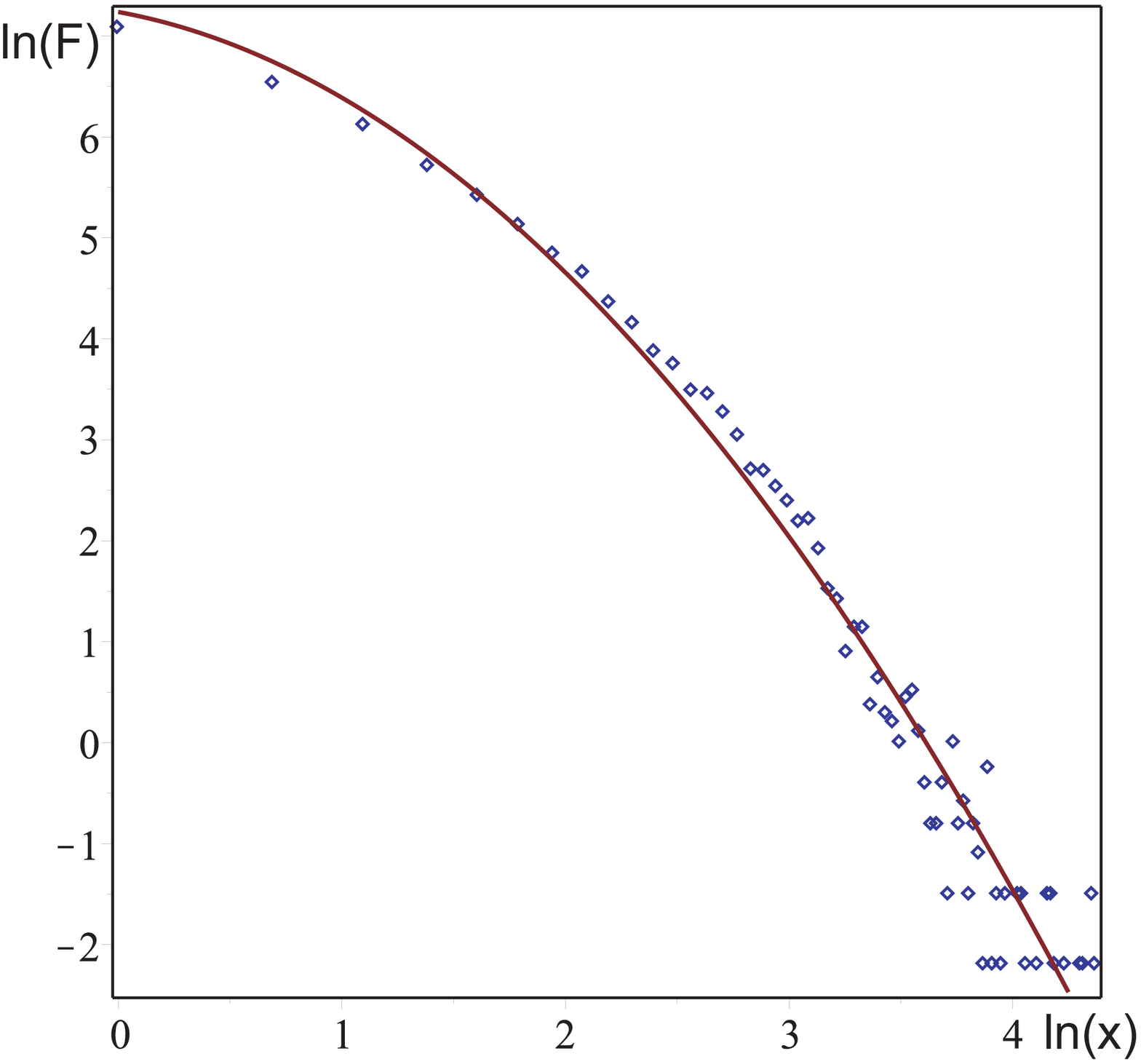}
    \includegraphics[width=5 cm]{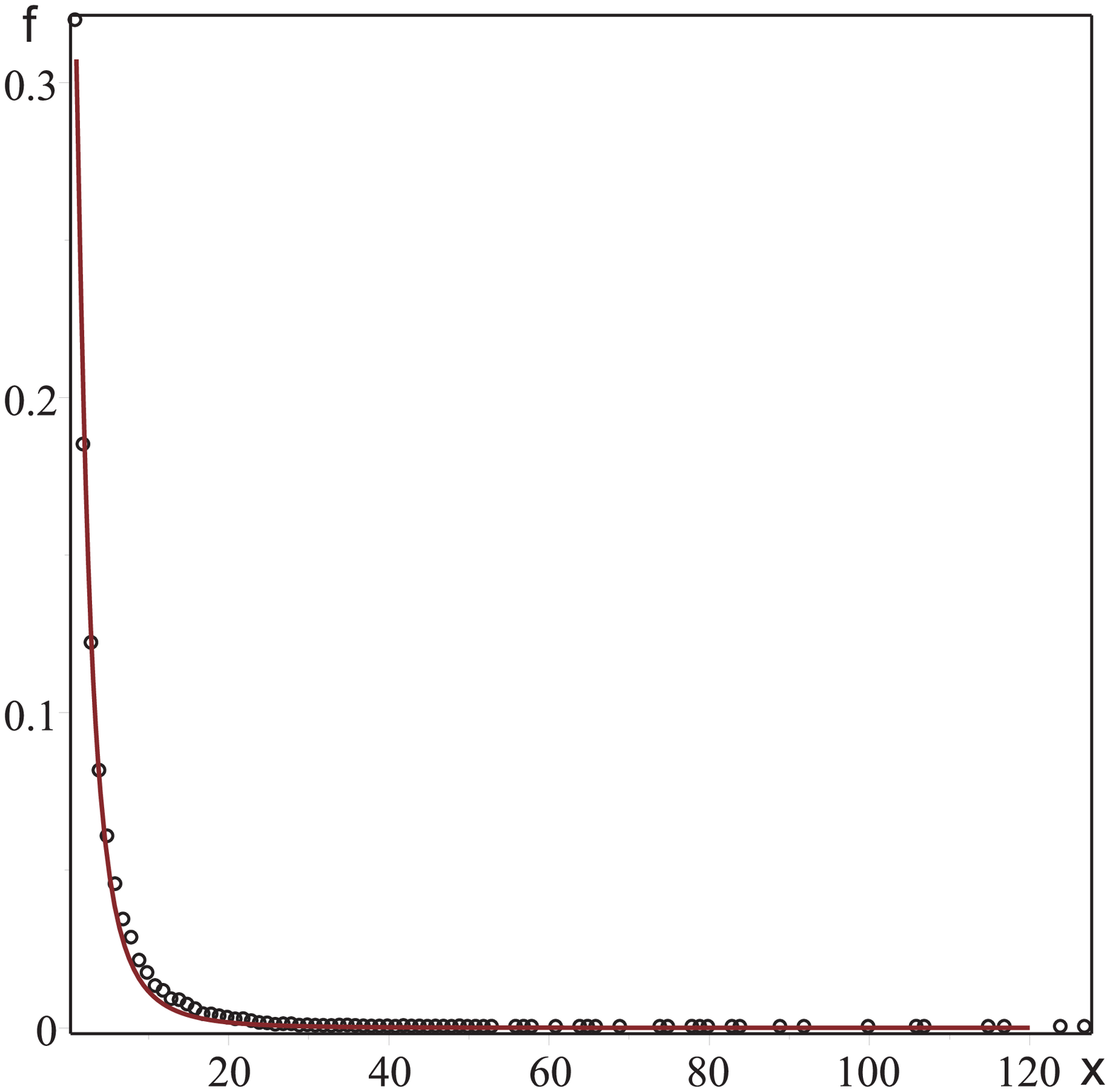}
  \caption{Distribution of incoming calls per day $x$ (left). Averaging was carried out over 9 days. Data in logarithmic coordinates (middle). The normalized distribution density of incoming calls is shown by circles (right). The continuous curve corresponds to the lognormal distribution density with the obtained parameters and from the approximating dependence.}\label{fg2}
\end{figure}

We now turn to a discussion of the density of incoming calls distribution over one day. Incoming calls of the same subset of subscribers and period were observed. The experimental data in the form of a histogram is shown in Fig.\ref{fg2}. Histogram averaging was also performed for 9 days. The total number of incoming calls according to the histogram is $N_{in}=3706$.

The quadratic approximation of the data, shown in Fig.\ref{fg2} in the middle, by the least-squares method leads to dependence  $\ln (y)=7.24-0.41 \ln (x) - 0.442 \ln (x)^2$. As above, the known values of the coefficients allows us to set the values of the parameters of the lognormal distribution density for incoming calls: $\mu \approx 0.67$ and $\sigma^{2} \approx 1.13$. Fig.\ref{fg2} shows the experimental values normalized to $N_{in}=3706$ and the continuous curve corresponding to the lognormal distribution density for the found values of $\mu$ and $\sigma$. A good agreement between these dependencies is observed. It is possible to achieve better agreement, given that the normalization of the experimental points was taken at an underestimated value. The reason for this is insufficient statistics and the presence of a large number of zeros on the tail of the distribution function.

Thus, the statistics of incoming calls also leads to a lognormal distribution with parameters $\mu \approx 0.7$ and $\sigma^{2} \approx 1.1$. Average value of incoming calls is $\mu^{\ast}\approx 3.4$ and variance is $\sigma^{\ast 2} \approx 6.7$ according to the lognormal law. Comparing the distribution of outgoing and incoming calls, you can notice that the average value of outgoing calls $\mu^{\ast}\approx 8$ significantly exceeds the average value of incoming calls $\mu^{\ast}\approx 3.4$. This means that the set of subscribers who called during the day is less than the set of all subscribers. Using the lognormal distribution function, in principle, this number can be calculated. Moreover, using the total experimental data for 9 days, the number of subscribers who called 1 time in 9 days can be determined.

\section{Spammers calls distribution}

\begin{figure}
  \centering
  \includegraphics[width=5 cm]{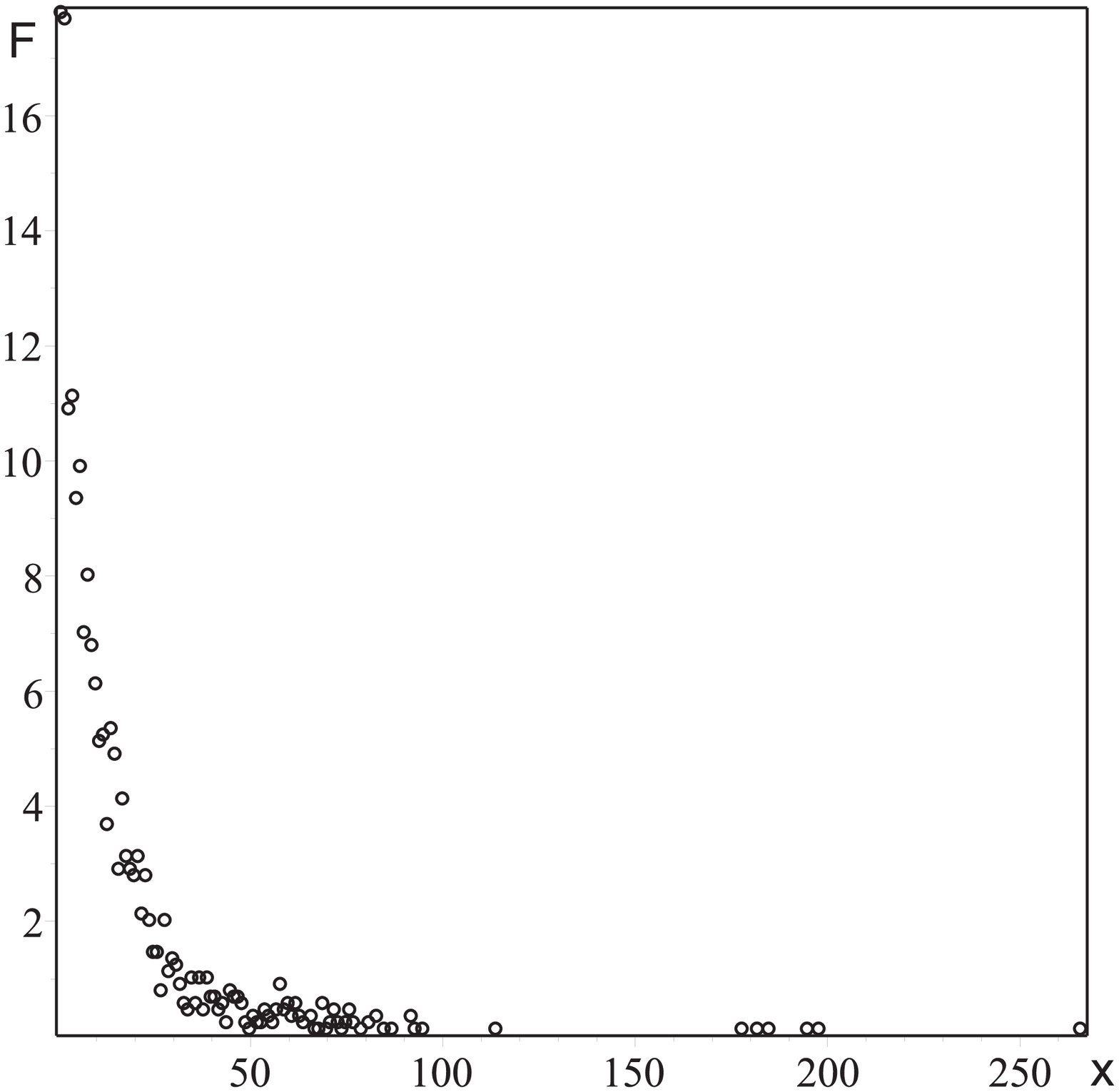}
   \includegraphics[width=5 cm]{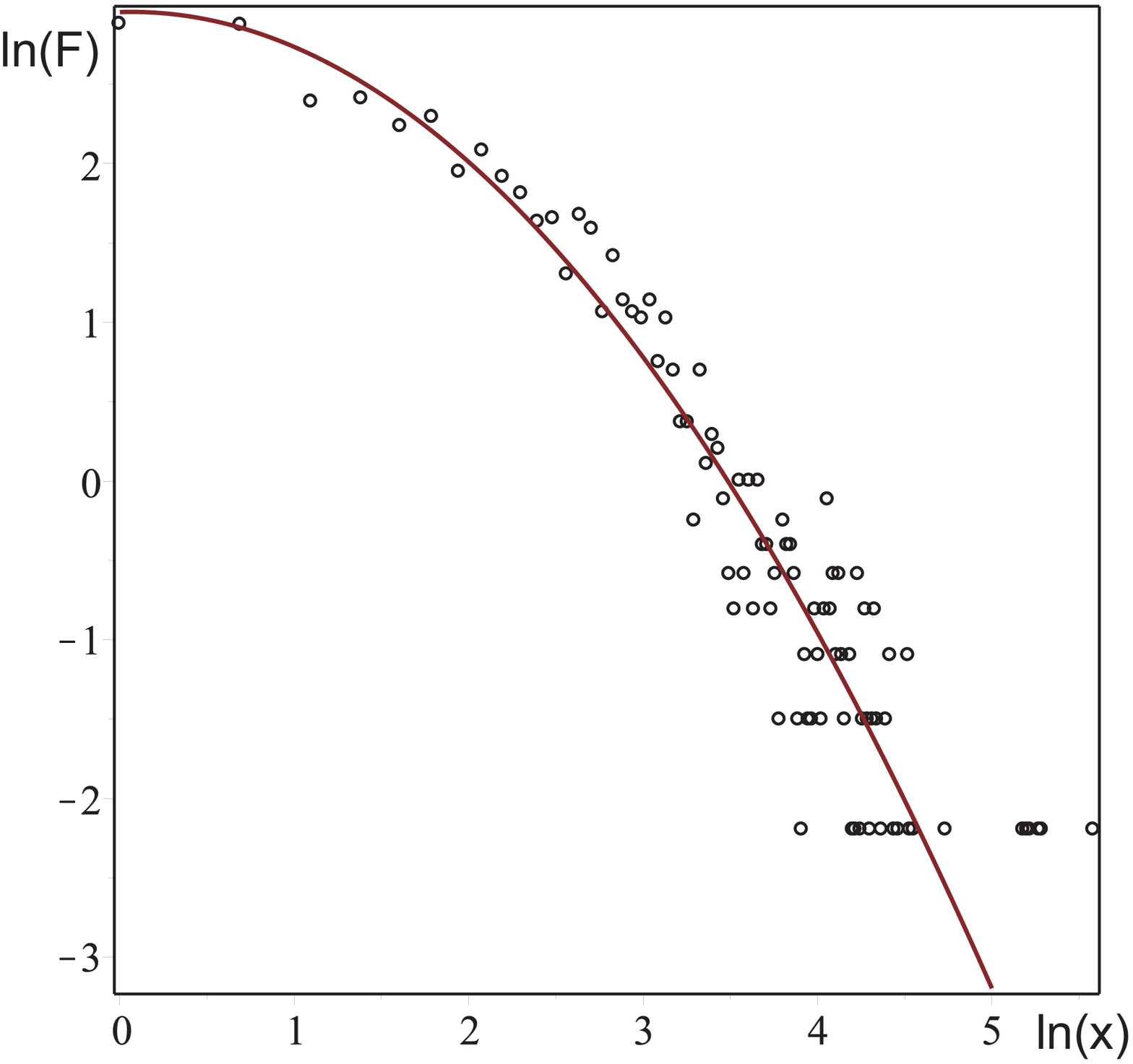}
   \includegraphics[width=5  cm]{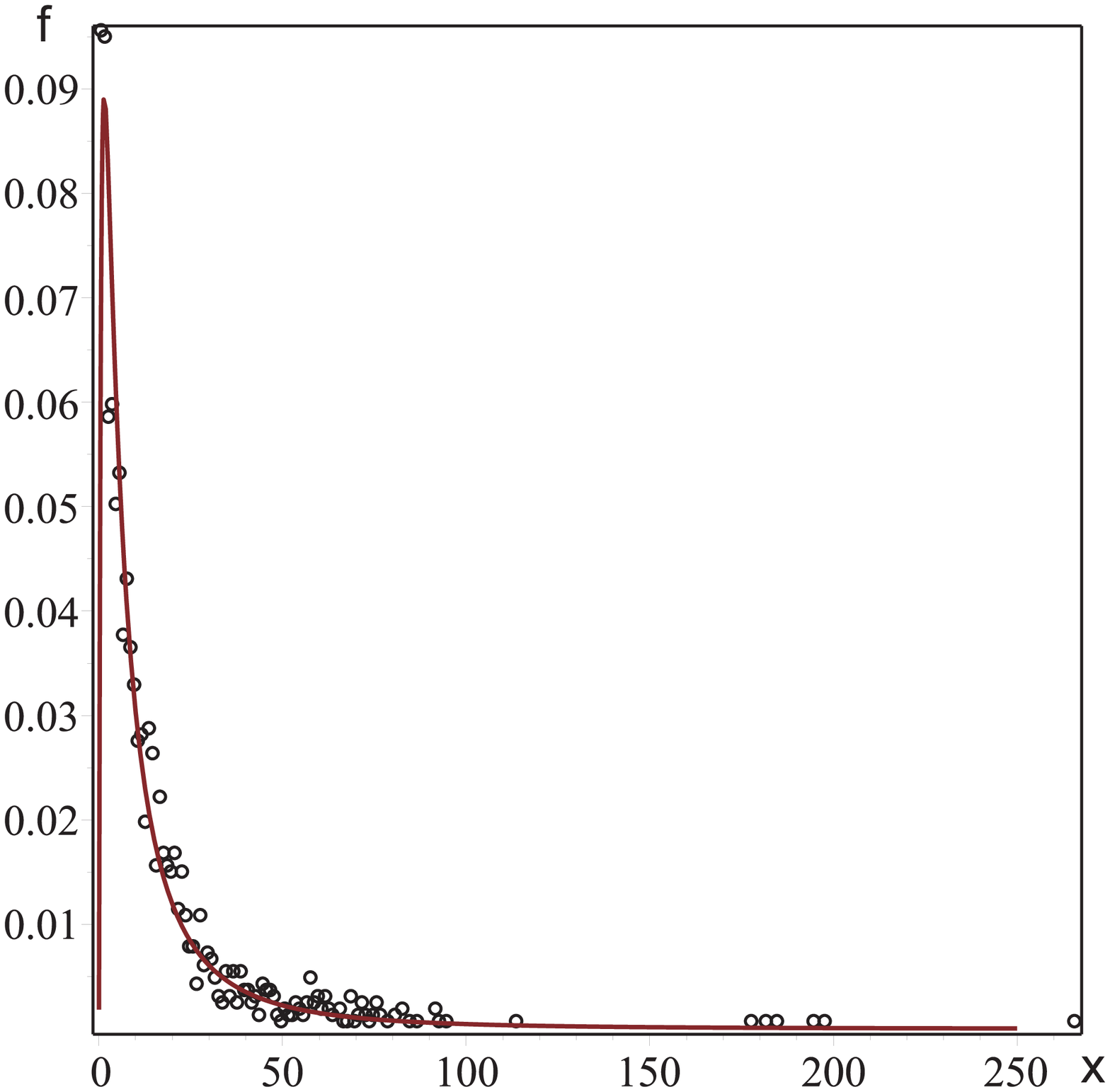}
     \caption{Distribution of incoming calls per day from spammers (left). Averaging was carried out over 9 days. Data in logarithmic coordinates (middle). A continuous curve is the dependence obtained using the least squares method. The normalized distribution density of incoming calls of spammers is shown by circles (right). The continuous curve corresponds to the lognormal distribution density with the obtained parameters $\mu$ and $\sigma$ from the approximating dependence.}\label{fg3}
\end{figure}
\begin{figure}
  \centering
  \includegraphics[width=5 cm]{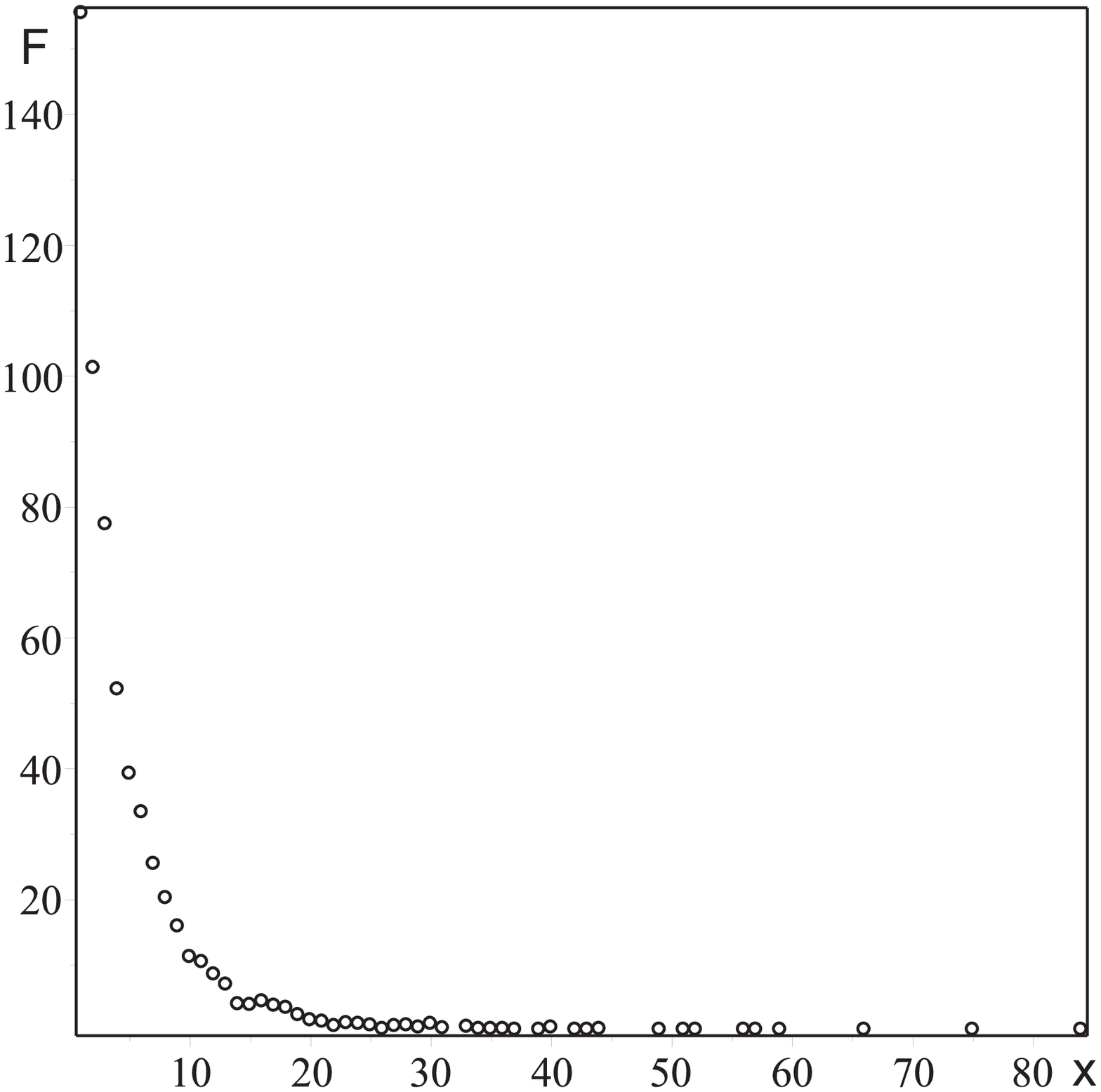}
   \includegraphics[width=5 cm]{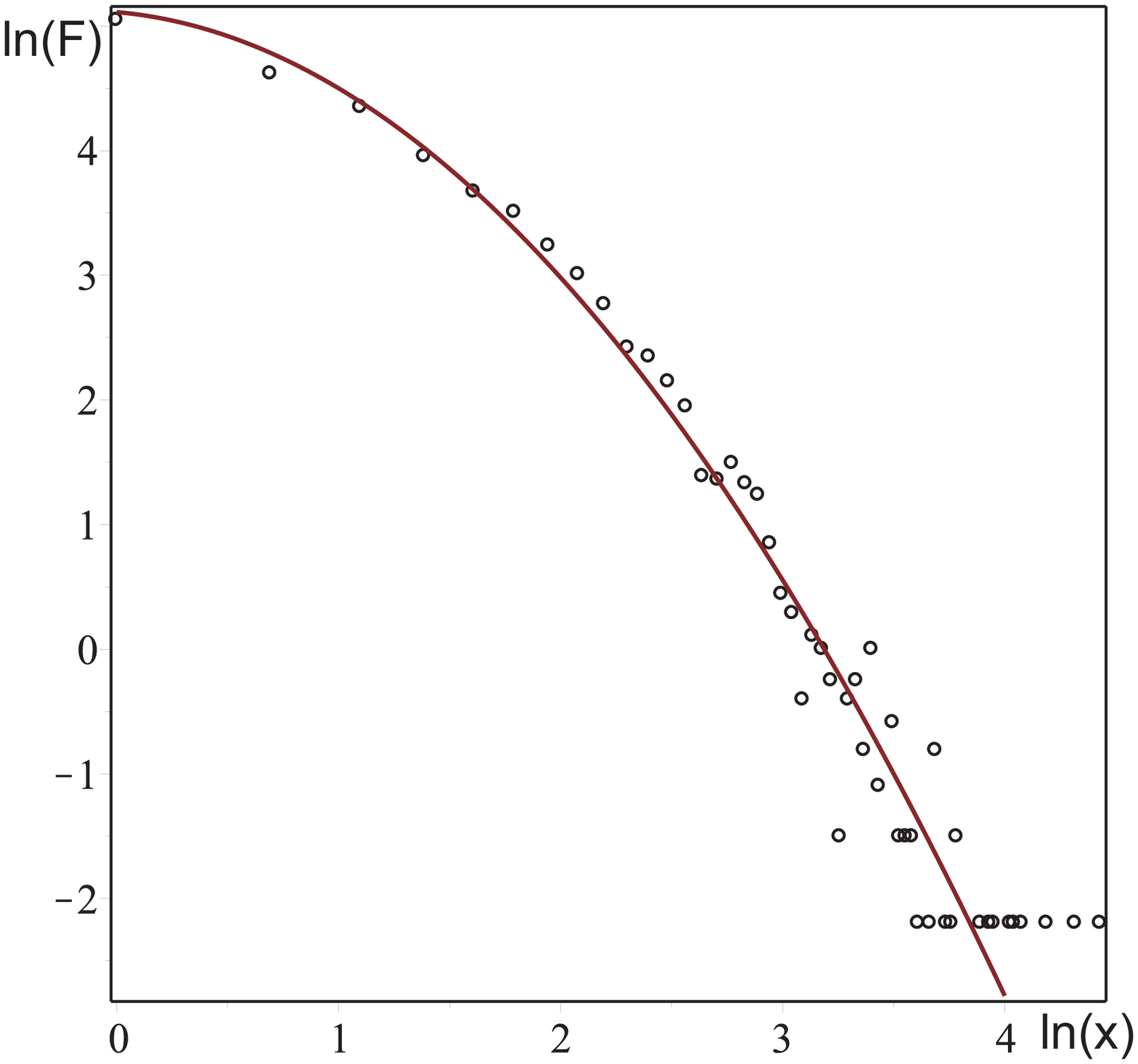}
    \includegraphics[width=5 cm]{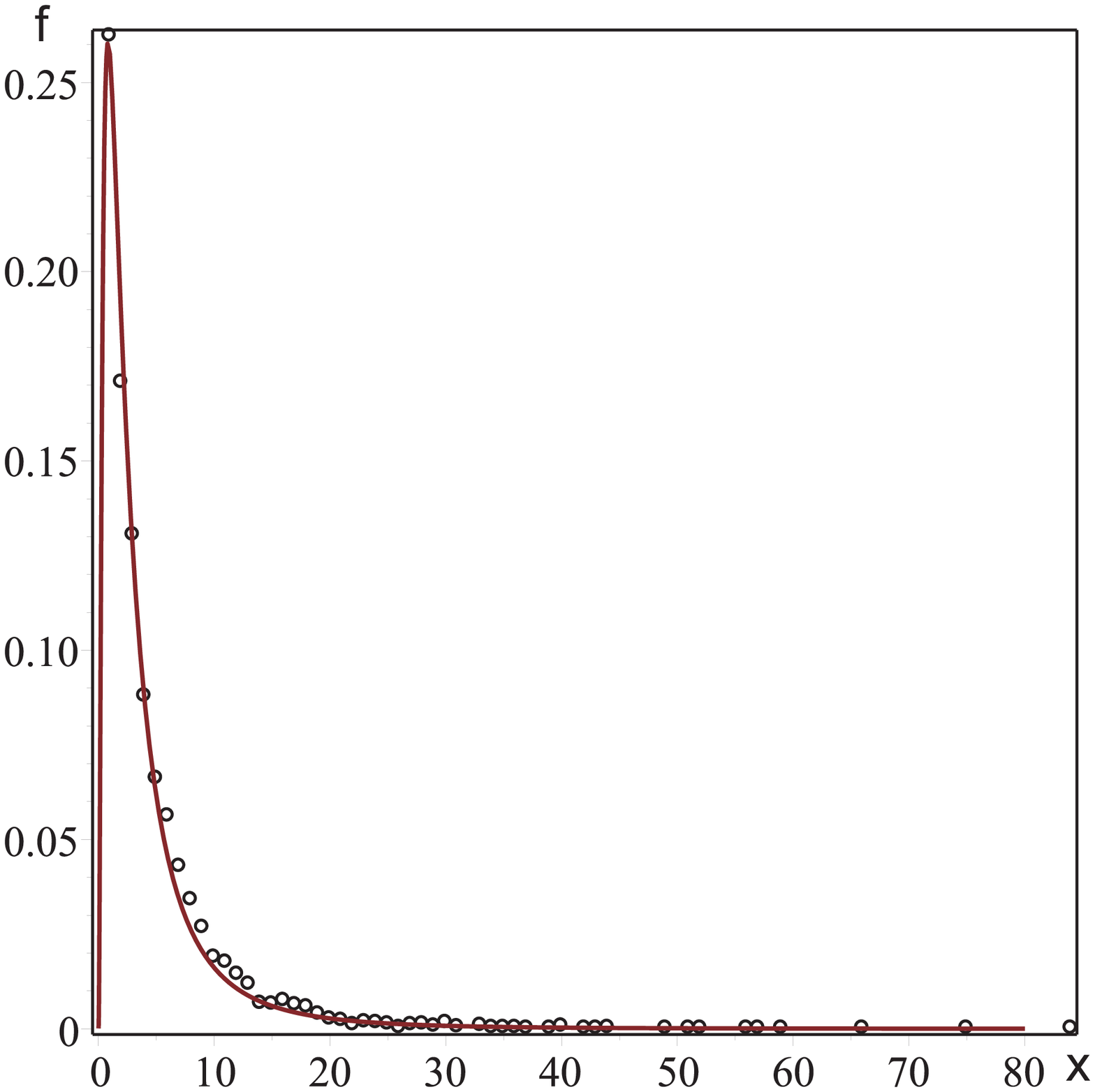}
  \caption{Distribution of outgoing calls per day from active subscribers (left). Averaging was carried out over 9 days. Data in logarithmic coordinates (middle). A continuous curve is the dependence obtained using the least squares method. The normalized distribution density of outgoing calls from active subscribers is shown by circles (right). The continuous curve corresponds to the lognormal distribution density with the obtained parameters $\sigma$ and $\mu$  from the approximating dependence.}\label{fg4}
\end{figure}

Consider now two more types of subscribers. Some of them can be called spammers, who annoy others with their calls, advertisements, etc. The second type of subscriber is active, who report the numbers of spammers with the requirements of their blocking. Let's start with a discussion of statistics about calls by spammers, based on information about them coming from active subscribers. Averaged statistics on spammers are shown in the histogram in Fig.\ref{fg3} on the left. The observation and averaging period are the same. The same data on a double logarithmic scale is shown in Fig.\ref{fg3} in the middle. The continuous curve shows the approximation dependence that was obtained by the least-squares method, which corresponds to a quadratic dependence
\[\ln f = 2.95+0.33 \cdot \ln(x)-0.25 \cdot \ln(x)^2\]
Using this relation, we can verify that the density of the distribution function corresponds to the lognormal distribution density with the parameters $\mu \approx 2$ and  $\sigma^{2} \approx 3$. The normalization factor for obtaining the distribution density over the histogram is $N_s =186$. Using this distribution density, the average value $\mu^{\ast} \approx 18$ and variance $\sigma^{\ast 2} \approx 146$  of spammer calls are calculated. It is easy to notice a significant excess of the average value of outgoing calls from spammers compared to the average value of outgoing calls from normal subscribers. Also, a large variance value means that individual subscribers can receive significantly more calls than the average value. The consistency of the experimental data and the lognormal distribution density is shown in Fig.\ref{fg3}. You can see a good agreement between these dependencies.
We now turn to the data about active subscribers. Let's start with statistics about outgoing calls of this subscriber group. The corresponding histogram is shown in Fig.\ref{fg4} on the left and on a double logarithmic scale in the middle. Quadratic dependence is observed. A similar approximation leads to a dependence of the form
\[\ln f = 5.11-0.161\cdot \ln(x)-0.45 \cdot \ln(x)^2\]
As before, the coefficients of this dependence determine the parameters of the lognormal distribution density. So for outgoing calls $\mu \approx 0.93$  and $\sigma^{2} \approx 5$. The values of these parameters are close to the values for all observed subscribers. The average values of the lognormal distribution density are $\mu^{\ast} \approx 4$  and $\sigma^{\ast 2} \approx 11$ also differ slightly, and $\mu^{\ast} \approx 8$ and $\sigma^{\ast 2} \approx 40$ for all observed subscribers. It is interesting to note that, on average, active subscribers call less frequently and the deviation from the average is less. It can be assumed that this is what makes them "sensitive" to calls from spammers.

Fig.\ref{fg4} shows the distribution density constructed from experimental data (circles) and a continuous curve of the lognormal distribution density. A good agreement of dependencies is seen.

\begin{figure}
  \centering
  \includegraphics[width=5 cm]{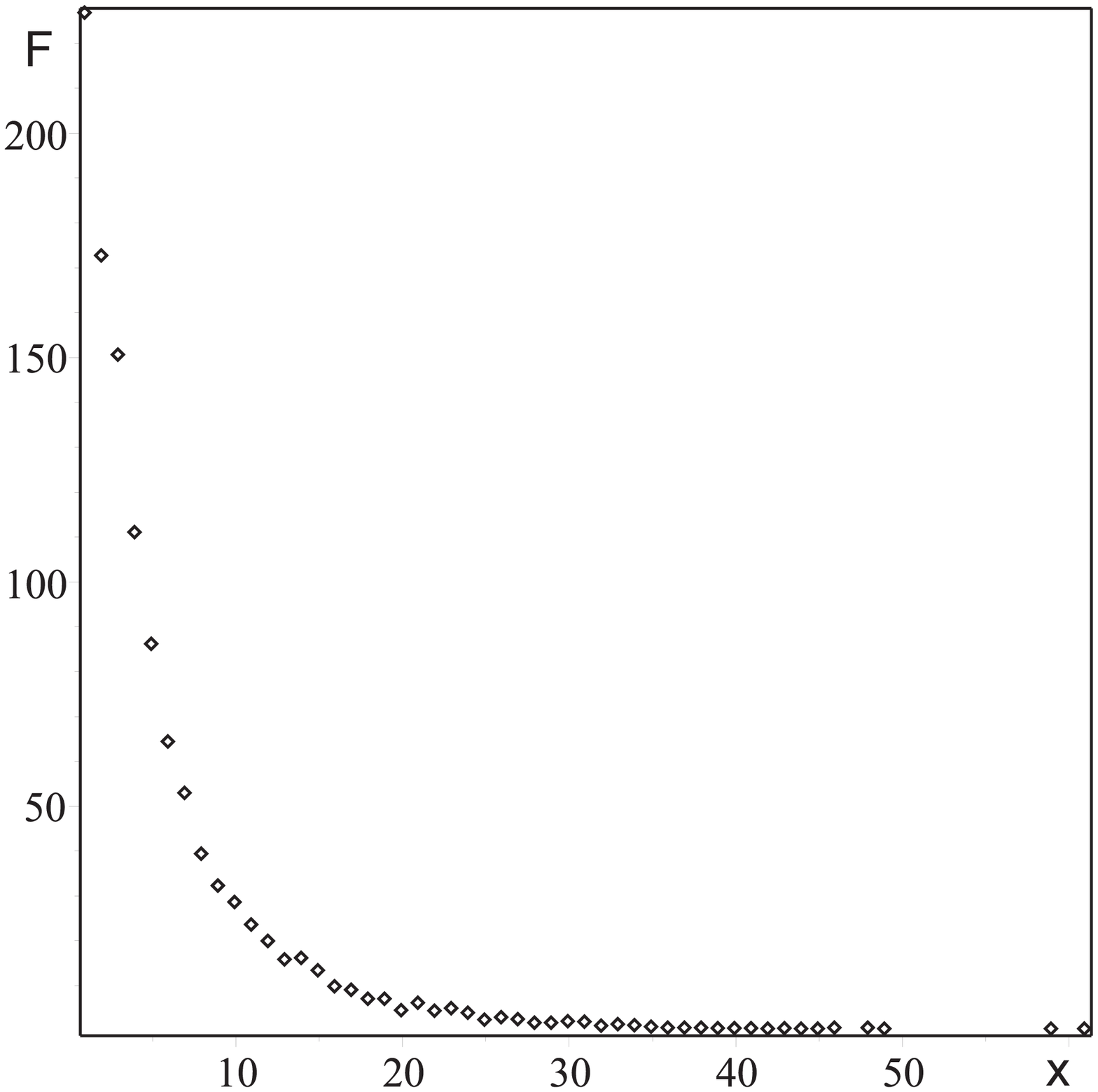}
   \includegraphics[width=5 cm]{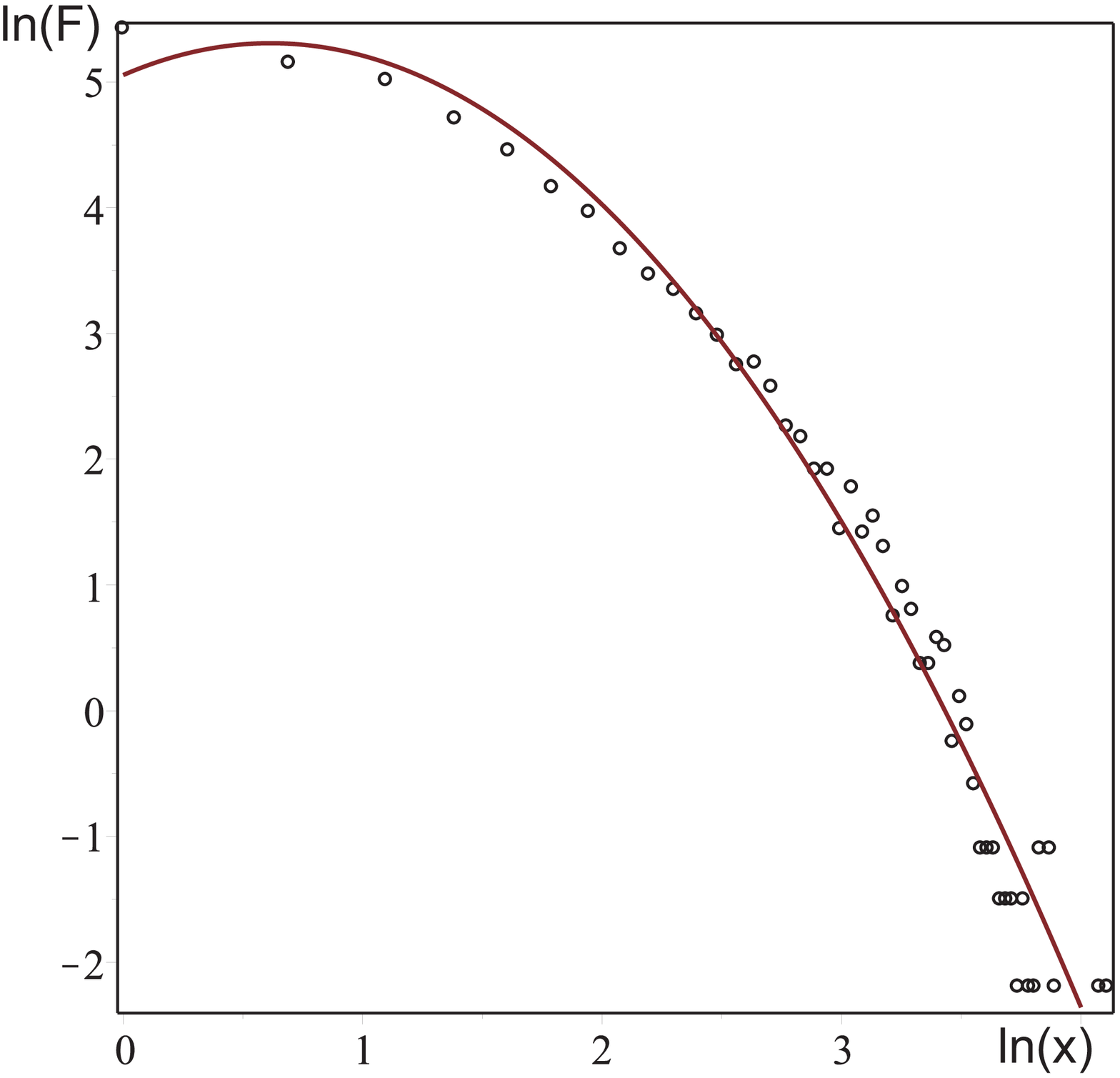}
   \includegraphics[width=5 cm]{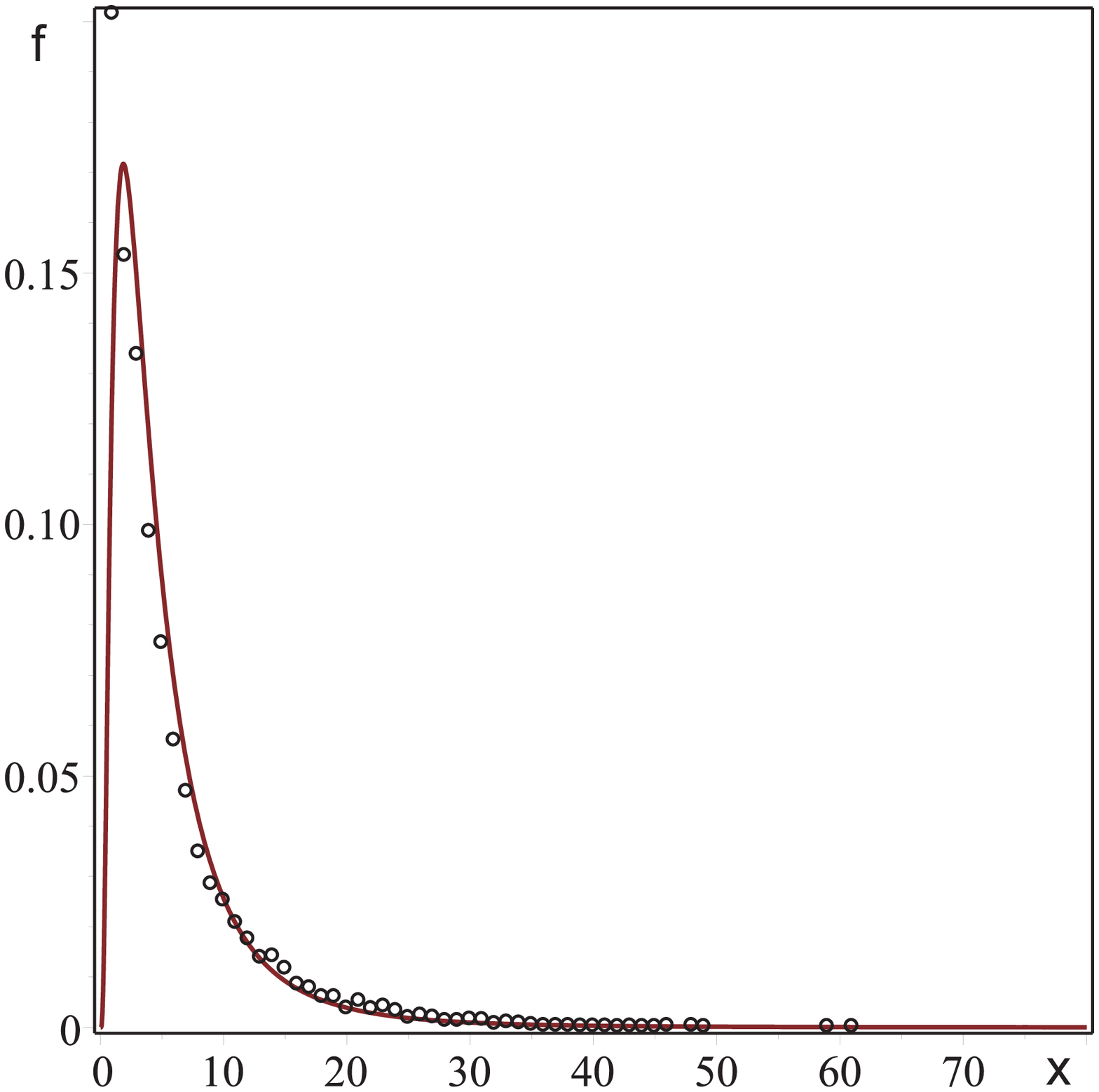}
  \caption{Distribution of incoming calls per day from active subscribers (left). Averaging was carried out over 9 days. Data in logarithmic coordinates (middle). A continuous curve is the dependence obtained using the least squares method. The normalized distribution density of incoming calls to active subscribers is shown by circles (right). The continuous curve corresponds to the lognormal distribution density with the obtained parameters  and   from the approximating dependence.}\label{fg5}
\end{figure}

Now we discuss the statistics of incoming calls to active subscribers. Fig.\ref{fg5} shows the corresponding data in usual and logarithmic coordinates. Based on these data, we again find the parameters of the lognormal distribution density using approximation of experimental data $\ln f = 5.06+0.82 \cdot \ln(x)-0.67 \cdot \ln(x)^2$. For statistics of incoming calls, the parameters take values $\mu \approx 1.36$, $\sigma^{2} \approx 5$. As before, this leads to an average number of incoming calls $\mu^{\ast} \approx 6$  and variance $\sigma^{\ast 2} \approx 22$. These values exceed the corresponding values for normal subscribers. Consequently, active subscribers do receive more incoming calls

Thus, the statistical properties of telephone network subscribers are determined by the lognormal distribution density, the parameters of which depend on the method of choosing communication type and subscribers subset. The reason why the lognormal distribution is realized is probably related to the mechanism that was discovered long ago in \cite{3s}. We reformulate them in application to phone calls. Let the number of calls  be determined in discrete time by the following equation
\[N_{i+1} - N_i = \omega_i N_i\]
$\omega$ - random frequency of calls. In other words, the rate of change in the number of calls is proportional to the random frequency and the number of calls made before. Solving this equation with the initial condition $N_0=1$ is easy to obtain
\[N_i =\prod_{j=0}^i (1-\omega_j)\]
After calculating the logarithm of this solution, we obtain the equality
\[\ln N_i = \sum_{j=0}^i \ln(1+\omega_j)\]
Note that random frequencies should be small. Otherwise, we would call very often. Then, given the small frequency, we restrict ourselves to the first term of the expansion of the logarithm and obtain
\[ln N_i = \sum_{j=0}^i \omega_j\]
Thus, the statistical properties of the logarithm of the number of calls are determined by the statistical properties of the sums of small random variables. Then we can use the central limit theorem on the distribution of sums of independent random variables (see, for example, \cite{4s}). According to which the sums of a sufficiently large number of independent quantities having finite average values and variance are normally distributed. This means, in our case, the normality of the distribution of the logarithms, or that the distribution density of the number of calls is lognormal.

\section{Discussion of the results}

Thus, the number of connections, both incoming and outgoing, are lognormally distributed. The difference lies in the valuesof the parameters $\mu$ and $\sigma$, which determine these distributions. For incoming and outgoing connections, the differences between these parameters are small. So, for example, the parameters of the lognormal distribution of outgoing calls are $\mu_{out} \approx 1$,  $\sigma^{2}_{out} \approx 1$, slightly differ from the parameters of incoming calls $\mu_{in} \approx 0.7$ $\sigma^2_{in} \approx 1.1$. However, the average values and variance of these lognormal distribution densities differ significantly. So for the above mentioned values ,  $\mu_{\ast out} \approx 8$, $\sigma^{2}_{\ast out} \approx 40$, and   $\mu_{\ast in} \approx 3.4$ $\sigma^2_{\ast in} \approx 6.7$. Therefore, to refine these parameters, even larger the corresponding amounts of experimental data should be used. In fact, these parameters determine the corresponding complex network of links.

It should be noted that although the telephone network of subscribers is discussed in the work, the main conclusion about the lognormality of vertices on degrees will remain valid for many other networks. For example, it can be expected for a network of firms, banks, etc., communicating with each other. All basic laws will befitted for them.

\end{document}